\documentclass[11pt]{article}
\newcommand{\bi}{\bibitem}
\newcommand{\be}{\begin{eqnarray}}
\newcommand{\ee}{\end{eqnarray}}

\begin{document}

\begin{center}{COSMIC ANTIMATTER: MODELS AND
PHENOMENOLOGY
}
\end{center}
\begin{center}
{ A.D. Dolgov }
\end{center}
\begin{center}
\it{ITEP, 117218, Moscow, Russia\\
{ { INFN, Ferrara 40100, Italy}}\\
{  University of Ferrara, Ferrara 40100, Italy}}
\end{center}


\begin{center}
Abstract
\end{center}
The possibility of creation of cosmologically significant antimatter
are analyzed in different scenarios of baryogenesis. It is argued that
there may exist plenty of antimatter even in our Galaxy. Possible forms
of antimatter objects and their observational signatures are discussed.
\\[5mm]


Prediction of antimatter made by P.A.M. Dirac~\cite{dirac} and the
discovery of anti-electron, i.e. positron~\cite{andersen} was surely
the most impressive result of quantum field theory of the XX century.
The whole new world of elementary particles was predicted and Dirac
naturally suggested that ``
If we accept the view of complete symmetry between positive and negative
electric charge so far as concerns the fundamental laws of Nature, 
we must regard it rather as an accident that the Earth 
(and presumably the whole solar system), contains a preponderance 
of negative electrons and positive protons. 
It is quite possible that for some of the stars it is the other way about, 
these stars being built up mainly of positrons 
and negative protons. In fact, there may be half of stars of each kind.''

It is striking that 30 years before Dirac and one year after discovery
of electron (J.J. Thomson, 1897)
A. Schuster~\cite{schuster} (another British physicist) conjectured that 
there might be other sign electricity, which he called {\it antimatter}, 
and supposed that there might be 
entire solar systems, made of antimatter and indistinguishable 
from ours. He even made an ingenious (or wild) guess that 
matter and antimatter are capable to annihilate and produce vast 
energy.

Today it is established beyond any doubt that for each particle there 
exists an antiparticle (except for a few absolutely neutral ones, as 
e.g. photons) but the expectations that half of the universe may consist 
of antimatter are more dead than alive. To a large extent this attitude 
is stimulated by the Sakharov theory of baryogenesis~\cite{sakharov} which
successfully explained the observed cosmological predominance of matter over 
antimatter. 

Though observations strongly restrict possible existence of antimatter domains
and objects, as is discussed below, it is still not excluded that antimatter
may be abundant in the universe and even in the Galaxy, not too far from us. 
This remaining freedom will be either eliminated by the 
existing (BESS~\cite{bess}, AMS~\cite{ams}, PAMELA~\cite{pamela})
and planned (AMS-2~\cite{ams-2}, PEBS~\cite{pebs}, and GAPS~\cite{gaps})
missions for search of cosmic antimatter or 
anti-stars or total anti-solar systems, 
as was envisaged by Schuster and Dirac, will be discovered.

Possible existence of anti-worlds depends upon the mechanism of
breaking of symmetry between particles and antiparticles, 
i.e. on physics of C and CP violation. From the cosmological point of
view there are three types of such violation:\\
1. Explicit, by complex parameters in Lagrangian, as is usually assumed  
in particle physics. \\
2. Spontaneous, by a vacuum expectation value (v.e.v.) 
of a real pseudoscalar field~\cite{lee-74} or a complex scalar 
or pseudoscalar field. Such mechanism is locally indistinguishable from the
explicit one but globally leads to charge symmetric universe, consisting of equal 
amount of matter and antimatter. This mechanism suffers from a serious domain
wall problem which jeopardizes homogeneity and isotropy of the observed
universe~\cite{zko}.\\
3. Dynamical or stochastic CP-breaking~\cite{dyn-cp}
enforced by a (pseudo)scalar field shifted from 
the equilibrium position and not
yet relaxed to equilibrium point at baryogenesis.
It could operate only in the early universe and disappeared without trace 
today and thus it is free of the domain wall problem. Of course an additional
source of CP-violation effective only in cosmology is
at odds with the Occam razor, but nevertheless it must 
work in the early universe if there exists a scalar field with mass smaller
than the Hubble parameter during inflation, $m<H_{inf}$.

Scenarios of baryogenesis with explicit C(CP)-violation lead to creation
of the universe devoid of noticeable antimatter, at least in 
the simplest versions of the models. 

Spontaneous C(CP)-violation evidently predicts alternating matter
and antimatter domains but, according to the results of ref.~\cite{c-dr-g} the 
domain size should be very large so that the nearest domain may be at least at the 
distance comparable to the cosmological horizon, $l_B \sim $ Gpc. However, in some
versions of the scenarios of the domain wall formation and elimination this bound
may be relaxed (work in progress).

An interesting possibility is the combined action of explicit plus spontaneous
C(CP)-violation~\cite{spont-expl} 
but probably it is strongly restricted, if not excluded, because 
it leads to too large
angular fluctuations of CMB and distortion of big bang nucleosynthesis.

The third mechanism of stochastic C(CP) breaking is probably most suitable 
for creation of cosmologically interesting antimatter, moreover such mechanism
naturally works in popular spontaneous~\cite{spont-bs} and Affleck and 
Dine~\cite{ad-bs} scenarios of baryogenesis. However, the versions of 
the models of cosmic antimatter creation suggested   
ref.~\cite{khlop} seem to suffer from excessive density 
perturbations at large scales which are excluded by observations.

A promising scenario of significant antimatter creation in our
neighborhood was proposed in ref.~\cite{ad-js}, which I will discuss in some
details below, as well as its cosmological manifestations~\cite{cb-ad}. 
The model predicts that antimatter is mostly concentrated in compact
stellar like objects (probably dead stars)
or dense clouds of antimatter and usual matter. 
These dense matter and antimatter objects have equal share in the total
cosmological mass density and
may contribute significant fraction of dark matter in our galaxy and in the 
universe, possibly even 100\%.

On the other hand, there are quite strong
observational bounds on possible antimatter in the universe.
The nearest antigalaxy in our supercluster cannot be closer than 
at $\sim$10 Mpc~\cite{steigman-76}. Otherwise the annihilation of intergalactic
gas from our supercluster inside such an anti-galaxy would create too large 
flux of gamma quanta with energy around 100 MeV.
Analogous arguments exclude fraction of anti-matter inside two colliding
galaxies (Bullet Cluster) larger than $ 3\times 10^{-6}$~\cite{steigman-08}.
These bounds together with cosmic ray data, that the 
fraction of ${\bar p}$ in cosmic rays is less than ${10^{-4}}$ and the
fraction of antihelium is smaller than  ${3\times 10^{-8}}$~\cite{bess} seem
to exclude large amount of antimatter in our Galaxy. However these bounds are
valid if antimatter makes exactly the same type objects as the OBSERVED matter,
as is usually assumed. 
For example, compact objects made of antimatter may be abundant,
live in the Galaxy, but still escape observations, as wee in what follows.

To create compact high density baryonic and antibaryonic objects we
rely on the Affleck-Dine mechanism of baryogenesis~\cite{ad-bs}. According 
to this mechanism a scalar baryon, $\chi$, condenses
along flat directions of its potential and accumulates very big baryonic
charge, $B$, later released in the decays of $\chi$ into quarks.
If $\chi$ acquires high $B$ value homogeneously in all space, it would end up
with the universe with constant and large baryon or antibaryon asymmetry.
However, if the window to the flat direction is open only during 
a short period, cosmologically small but possibly astronomically large 
bubbles with high ${B}$ could be
created, occupying {a small fraction of the cosmological volume, 
while the rest of the universe would have the normal 
baryon asymmetry:
\be 
\beta = N_B/N_\gamma \approx 6\times 10^{-10},
\label{beta}
\ee
while inside the high B-bubbles the asymmetry may be of order unity or even larger.

Such rather unusual inhomogeneous modification of the Affleck-Dine mechanism can 
be realized by a very simple modification of the
potential of the Affleck-Dine field $\chi$.
We assume that $\chi$ has the usual Coleman-Weinberg potential~\cite{cw-pot}
plus general renormalizable coupling to inflaton field $ \Phi$:
\be 
U(\chi,\Phi)  = {\lambda_1|\chi|^2 (\Phi -\Phi_1)^2} + 
\lambda_2 |\chi|^4 \,\ln (|\chi|^2/\sigma^2) +
m_0^2 |\chi^2| + (m^2_1 \chi^2 + h.c.),
\label{U-of-chi}
\ee
where $\Phi_1$ is some constant value which $\Phi$ passes in the course of
inflation but not too long before the end of inflation, otherwise the high B-regions
would be too large. 
The mass parameter $ m_1$ may be complex but CP would be still conserved, 
because one can ``phase rotate'' $ \chi$ to eliminate complex parameters 
in the Lagrangian. It is essential that the last term is not invariant
with respect to $U(1)$ transformation, $\chi \rightarrow \exp(i\Theta) \,\chi$, 
and thus it breaks B-conservation.
Potential (\ref{U-of-chi}) has one minimum at $\chi =0$ for large and 
small $\Phi$ and has a deeper minimum at non-zero $\chi$ when $\Phi$ 
is close to $\Phi_1$. At that time the chances for $\chi$ to reach a high
value at the other minimum are non-negligible.

The probability for $\chi$ to reach the deeper minimum is determined by the
quantum diffusion at inflationary stage. It is governed 
by the diffusion equation~\cite{starobinsky}:
\be 
\frac{\partial{\cal P}}{\partial t}=
\frac{H^3}{8\pi^2}\sum_{k=1,2}\frac{\partial^2{\cal P}}{\partial\chi_k^2}+
\frac{1}{3H}\sum_{k=1,2}\frac{\partial}{\partial\chi_k}
\left[{\cal P}\frac{\partial U} {\partial\chi_k}\right]
\label{dP-dt}
\ee
where ${\chi= \chi_1+i\chi_2}$.
The inflation may be not exact and $H_I$ may depend upon time but
this does not significantly influence the spectrum of the produced bubbles with
high baryonic density.

It can be shown~\cite{ad-js} that the mass spectrum of the high $B$ bubbles has 
log normal form:
\be
\frac{dn}{dM} = C_M \exp{[-\gamma \ln^2 (M/M_0)]}
\label{dn-dM}
\ee
where ${C_M}$, ${\gamma}$, and ${M_0}$ are some constant model dependent parameters.

According to this scenario 
the universe would look as a huge piece of swiss cheese with high $B$ bubbles
instead of holes and homogeneous background with constant baryonic density.
The mass of those high B objects is model dependent and can be
of the order of stellar mass or even larger or much smaller.
Despite their small size
the mass fraction of the bubbles could be comparable or even larger than
the observed baryonic mass fraction. 

Initially the density contrast between the bubbles with high values of $\chi$
and the bulk with $\chi \sim 0$ was small.
Later, when the mass of $\chi$ came into play, the matter inside
the bubbles with a large amplitude of $\chi$ became nonrelativistic and the
density contrast started to rise. The rise continued till $\chi$ decayed into
light quarks and/or leptons and the matter inside became relativistic as in the
bulk of the universe.
The second period of the rising perturbations took place after
the QCD phase transition at $T=T_{QCD} \sim 100$ MeV,
when relativistic quarks confined to make nonrelativistic nucleons.

The final state of the high B-bubble depends upon their mass and radius.
If the density contrast was large, 
${\delta\rho /\rho \geq 1}$ at horizon crossing, the bubbles would form
primordial black holes (PBH). The mass inside horizon at cosmological time $t$
is equal to:
\be
M_{hor} \approx  m_{Pl}^2 t\approx 10^{38} \rm{ g}\, ({ t}/\rm{ sec})
\approx 10^{5} M_\odot (t/sec),
\label{M-hor}
\ee
where $M_\odot$ is the Solar mass.
At the moment of QCD phase transition the mass inside horizon is about $M_\odot$, 
while during big bang nucleosynthesis (BBN) the mass inside horizon varies from
$10^5 M_\odot$ to to ${10^{7} M_\odot}$. So as a by-product the considered scenario 
allows for formation of superheavy black holes whose origin is not well explained
by conventional mechanisms.

One should keep in mind that not only black holes but some other
compact objects could be formed too. If PBH had not been formed, 
the subsequent evolution of the B-bubble depends 
upon the relation between their mass and the Jeans mass.
Bubbles with ${{ {\delta\rho}/{\rho}<1}}$ but with
${{M_B>M_{Jeans}}} $
at horizon {would decouple from cosmological expansion} and
form compact stellar type objects or lower density clouds.
Such anti-objects could survive against an early annihilation.
Moreover, the annihilation products would be practically unnoticeable.  
According to the estimates of ref.~\cite{cb-ad}, the fraction of extra energy
produced by the annihilation on the surface of the compact stellar-like 
anti-objects would not exceed about $10^{-15}$ of CMB. 

The bubbles with high baryonic number density, after they decoupled from the
cosmological expansion, would have higher temperature than the background.
The luminosity due to thermal emission into external space would be
${L\approx 10^{39}}$ erg/sec for the bubbles with solar mass.
Even if these high-B bubbles make all cosmological
dark matter, i.e. ${ \Omega_{BB} = 0.25}$, the thermal keV photons would 
have the energy density not more than ${10^{-4}-10^{-5}}$ of 
the energy density of CMBR, red-shifted today to the 
background light~\cite{cb-ad}.

Some of those high density baryonic and antibaryonic objects might
be similar to red giants core. The dominant internal energy source of 
these objects would be the nuclear helium burning through the 
reaction ${3He^4 \rightarrow C^{12}}$, however with larger T by factor
${\sim 2.5}$ in comparison with red giants.
Since $ {L\sim T^{40}}$, the nuclear power exhaust
would be very fast, with life-time shorter than $10^9$ s.
The total energy outflux would be below ${10^{-4}}$ of CMBR.
It is unclear it the nuclear reactions could lead to 
the B-ball explosion and creation of solar mass anti-cloud.
Astrophysics of such early formed high density objects is not yet studied.

Thus sufficiently large compact anti-objects could survive in the early universe,
especially if they are PBHs.
A kind of early dense stars might be formed with 
initial pressure outside larger than that inside, which introduced an additional
stabilizing factor for their survival. 
Such ``stars'' would evolve quickly and, in
particular, could make early supernovae which would
enrich the universe with heavy (anti)nuclei
and re-ionize the universe.
Energy release from stellar like objects in the early
universe is small compared to the energy density of CMB.
The objects with high (anti)baryonic number are not dangerous for the standard
outcome of big bang nucleosynthesis since the volume of such B-bubbles is 
small. However, all above are the results of some simplified estimates and more 
rigorous calculations are necessary.

In the present day universe the antimatter bubbles created by the discussed
above mechanism may form all kinds of antimatter objects populating the Galaxy
or its halo. They could make dense clouds of antimatter, isolated antistars,
which are most probably evolved and not shining, anti-black holes, which are
indistinguishable from the black holes made of the usual matter, except for 
possible anti-atmosphere of non-collapsed primordial antimatter.

The observational signatures of such antimatter objects are almost evident.
They include the standard list of 100 MeV gamma background, excessive antiprotons, 
and low energy positrons. For all these observations an alternative explanations
cannot be a priori excluded, but an observation of cosmic antinuclei, starting from
anti-helium and heavier ones would be unambiguous proof of existence of
primordial antimatter.

One can also expect to see compact sources of ${\gamma}$ radiation, which are
not associated with shining stars. The chemical content of those objects
maybe be different from the normal stars, because they were created from matter
processed through BBN with unusually high baryon-to-photon ratio. 
So an observation of compact stellar-like
objects in the sky or clouds with unusual chemistry would be a strong indication
that such objects are made of antimatter and search for annihilation around
them has chance to be successful.

There could also be some catastrophic phenomena generated by the star-antistar 
collisions, processes similar to hypernova explosions induced by transfer of material
in double star-antistar system. A capture of compact antistar by a red giant
would lead to a noticeable change of the red giant luminosity over a time interval
of the order of a month. 

To summarize both theory and observations allow for abundant cosmic antimatter
in the Galaxy and in its halo and the search of cosmic antinuclei, 
excessive antiprotons
and positrons, gamma-quanta from the antiproton and positron annihilation, and
striking astrophysical phenomena due to star-antistar interactions have a
non-negligible chance to be discovered.


\end{document}